\documentclass[aps,twocolumn,showpacs,superscriptaddress]{revtex4}
\usepackage{amsmath}
\usepackage{amssymb}
\usepackage{epsfig}
\usepackage{color}
\usepackage[colorlinks]{hyperref}
\hypersetup{colorlinks,citecolor=red,linkcolor=blue,urlcolor=blue}
\usepackage{graphicx,amsmath}
\begin{document}

\title{Common behavior of the scaled condensation energy for both
high-$T_c$ and conventional superconductors}
\author{V. R. Shaginyan}\email{vrshag@thd.pnpi.spb.ru}
\affiliation{Petersburg Nuclear Physics Institute, NRC Kurchatov
Institute, Gatchina, 188300, Russia}\affiliation{Department of
Physics, Clark Atlanta University, Atlanta, GA 30314,
USA}\author{A. Z. Msezane} \affiliation{Department of Physics,
Clark Atlanta University, Atlanta, GA 30314, USA}
\author{S. A. Artamonov} \affiliation{Petersburg Nuclear Physics Institute,
NRC Kurchatov Institute, Gatchina, 188300, Russia}

\begin{abstract}
We analyze the scaling of the condensation energy $E_{\Delta}$
divided by $\gamma$,  $E_{\Delta}/\gamma\simeq
N(0)\Delta_1^2/\gamma$, of both conventional superconductors and
unconventional high-$T_c$ one, where $N(0)$ is the density of
states, $\Delta_1$ is the maximum value of the superconducting gap
and $\gamma$ is the Sommerfeld coefficient. For the first time, we
show that the universal scaling of $E_{\Delta}/\gamma\propto T_c^2$
applies equally to conventional superconductors and unconventional
high-$T_c$ ones. Our consideration is based on both facts:
Bogoliubov quasiparticles act in conventional and unconventional
superconductors, and the corresponding flat band is deformed by the
unconventional superconducting state. As a result, our theoretical
observations based on the fermion condensation theory are in good
agreement with experimental facts.
\end{abstract}

\pacs {74.25.Bt; 74.72.-h; 64.70.Tg}

\maketitle

\section{Introduction}\label{I}

It is generally accepted that conventional superconductors have
nothing in common with unconventional superconductors, since
unconventional superconductors are strange metals with flat bands
\cite{catal} in the absence of quasiparticles, see e.g. \cite{scn}.
On the other hand, experimental facts show that both types of
superconductors have common properties, have quasiparticles and
exhibit common scaling behavior, see, for example,
\cite{prb2015,prlq,mat,npbq}; while the corresponding flat bands of
high-$T_c$ superconductors are deformed by superconducting state,
which makes high-$T_c$ superconductors similar to ordinary
superconductors \cite{mac,epl22}. Thus, these contradictions pose a
challenging puzzle for condensed matter researchers.

The flat band problem described above could have been solved many
years ago when Landau's Fermi liquid (LFL) theory was developed
\cite{lanl}. As known, it  deals with energy functionals
$E_0[n({\bf p})]$ in the functional space $[n]$ of quasiparticle
distributions $n({\bf p})$ located in $[n]$ between 0 and 1. This
theory is based on assumption that the single particle spectrum of
a normal Fermi liquid is similar to that of an ideal Fermi gas,
differing from the latter in the value of the effective mass $M^*$.
At temperature $T=0$, in a homogeneous isotropic matter, the LFL
ground state quasiparticle distribution is the Fermi step function
$n_F(p)=\theta (p-p_F)$. Quasiparticles fill the Fermi sphere up to
the same radius $p_F=(3\pi^2\rho)^{1/3}$ ($\rho$ is the number
density and $p_F$ is the Fermi momentum) as noninteracting
particles  do (the Landau-Luttinger theorem \cite{lanl}). From the
mathematical point of view, in the LFL, the minimum of $E_0[n]$ is
supposed to always lie at a boundary point $n_F$ of the space
$[n]$. This assumption remains valid as long as the necessary
stability condition
\begin{equation}\label{nsc}
\delta E_0=\int (\varepsilon[{\bf p},n( {\bf p},T=0)] -\mu)\delta
n({\bf p},T=0)\frac{d^3p}{(2\pi)^3}>0,
\end{equation}
is fulfilled. Here $\varepsilon[{\bf p},n({\bf p})]={\delta
E_0[n]/\delta n({\bf p})}$ is the quasiparticle energy, $n({\bf
p})$ is the quasiparticle distribution function, and $\mu$ is the
chemical potential. The stability condition requires that the
change of $E_0[n]$ for any admissible variations of $n_F$ holds.
Thus, it is the violation of the condition given by Eq. \eqref{nsc}
that results in the rearrangement of the distribution $n_F({\bf
p})$. The quasiparticle distribution function $n({\bf p})$ is
constrained by the Pauli principle $1\geq n({\bf p})\geq 0$. As a
result, there are two classes of solutions of Eq. \eqref{nsc}. One
class forming flat bands is
\begin{equation}\label{7**}
\varepsilon(p)=\mu;\,\, {\rm if}\,\, 1>n_0({\bf p})>0\,\, {\rm
in}\,\, p_i<p<p_f,
\end{equation}
which is valid if the special solution $n_0({\bf p})$ becomes
$1>n_0({\bf p})>0$ in some region $p_i<p_F<p_f$
\cite{ks,ksk,phys_rep,vol,Volovik}. The other conventional class is
defined by $\delta n({\bf p})=0$ with $n({\bf p})=0$ or $n({\bf
p})=1$, that is $n({\bf p})=n_F({\bf p})$ \cite{lanl}.

Flat bands, now observed in many strongly correlated Fermi systems
\cite{catal}, first emerged as a mathematical curiosity
\cite{ks,ksk} and now represent a rapidly expanding and dynamic
field with countless applications, see e.g
\cite{catal,book_20,phys_rep,Volovik,prl20,bern,bern_prl}.
High-$T_c$ superconductors represent a wide class of strongly
correlated Fermi systems, exhibiting the non-Fermi liquid (NFL)
behavior defined by flat bands, see e.g
\cite{phys_rep,Volovik,prl20,bern,bern_prl}. As a result, one can
expect that superconductors with high-$T_c$ have nothing in common
with conventional superconductors. For example, in case of
high-$T_c$ superconductors the critical temperature
\cite{ks,ksk,Volovik,prl20,bern,bern_prl}
\begin{equation}\label{TC}
T_c\propto \Delta_1\propto \lambda_0,
\end{equation}
rather than being $T_c\propto\exp{(-1/gN(0))}$, where $\lambda_0$
is the superconducting coupling constant, $\Delta_1$ is the maximum
value of the superconducting gap and $N(0)$ is the density of
states at the Fermi surface \cite{bcs,til}. However, in both
conventional and unconventional high-$T_c$ superconductors, the
condensation energy exhibits universal scaling behavior:
$E_{\Delta}/\gamma\simeq N(0)\Delta_1^2/\gamma\propto T_c^2$, as it
follows from experimental facts \cite{prb2015}.

In our paper we analyze both unconventional high-$T_c$
superconductors and conventional ones, and demonstrate that both of
them exhibit the common universal scaling of the condensation
energy $E_{\Delta}/\gamma$, $E_{\Delta}/\gamma\simeq
N(0)\Delta_1^2/\gamma$. For the first time, we explain  that the
universal scaling of $E_{\Delta}/\gamma\propto T_c^2$ applies
equally to conventional and unconventional high-$T_c$
superconductors. Our results are in good agreement with
experimental facts \cite{prb2015}. This observation suggests that
the FC superconducting state is BCS-like and suggests the
fundamental applicability of the BCS formalism to describe some
properties of the superconducting state, as predicted in
\cite{phys_rep,jetplbq}. Our analysis is made within the framework
of the fermion condensation (FC) theory based on the topological
fermion condensation quantum phase transition (FCQPT) that forms
flat bands and leads to the universal scaling behavior of the
thermodynamic and transport properties of HF metals
\cite{ks,ksk,phys_rep,book_20}.

\section{Superconducting systems with the FC state}\label{II}

Here we consider the superconducting state of high-$T_c$
superconductors within the framework of the FC theory
\cite{ks,phys_rep}. It was experimentally shown that in HF metals
the quasiparticles are well-defined excitations \cite{prlq} and in
the superconducting state of high-$T_c$ superconductors the
elementary excitations are Bogoliubov quasiparticle (BQ), that is
the excitations are Bardeen-Cooper-Schrieffer like
\cite{bcs,til,mat,npbq}. Therefore, as we shall see, unconventional
superconductors exhibit the same scaling behavior of the
condensation energy $E_{\Delta}/\gamma$ as conventional
superconductors do \cite{lanl}.

The energy dispersion of single-particle excitations and the
corresponding coherence factors as a function of momentum were
measured on high-$T_c$ cuprates
(Bi$_2$Sr$_2$Ca$_2$Cu$_3$O$_{10+\delta}$, $T_c$=108 K) by using
high-resolution angle-resolved photoemission spectroscopy
\cite{mat}. All the observed features qualitatively and
quantitatively agree with the behavior of BQ in conventional
superconductors predicted by the BCS theory
\cite{bcs,til,mat,prlq}. This observation shows that the
superconducting state of high-$T_c$ superconductors is BCS-like
with BQ, and implies the basic validity of the BCS formalism in
describing the superconducting state, and is closely related to the
deformation of flat band by the superconducting phase transition
\cite{shag,ms,Schuck,jetplbq,epl22}. On the other hand, a number of
the properties as the maximum value of the superconducting gap
$\Delta_1$, the high density of states and the other exotic
properties are beyond the BCS theory
\cite{phys_rep,jetplbq,book_20}.

Below we shall call electron (hole) liquids as electron one. At
$T<T_c$, the thermodynamic potential $\Omega$ of an electron liquid
is given by the Equation (see, e.g. \cite{lanl,til})
\begin{equation}\label{1*}
\Omega=E_{gs}-\mu N-TS,
\end{equation}
In Eq. \eqref{1*} $N$ is the number density of quasiparticles, $S$
denotes the entropy, and $\mu$ is the chemical potential. The
ground state energy $E_{gs}[\kappa({\bf p}),n({\bf p})]$ of
electron liquid is the exact functional of the order parameter of
the superconducting state $\kappa({\bf p})$ and of the
quasiparticle occupation numbers $n({\bf p})$
\cite{phys_rep,pla98}. Here we assume that the electron system is
two-dimensional in order to describe the results of Ref.
\cite{mat}, while all results can be transported to the case of
three-dimensional system. This energy is determined by the known
equation of the weak-coupling theory of superconductivity
\begin{equation}\label{2*} E_{gs}\ =\
E[n({\bf p})]+\delta E_s.
\end{equation}
Here $E[n({\bf p})]$ is the exact Landau functional determining the
ground-state energy of normal Fermi liquid \cite{lanl,phys_rep},
and $\delta E_s$ is given by
\begin{equation}\label{2**}
\delta E_s=\int\lambda_0V({\bf p}_1,{\bf p}_2)\kappa({\bf p}_1)
\kappa^*({\bf p}_2)\frac{d{\bf p}_1d{\bf p}_2}{(2\pi)^4}.
\end{equation}
Here $\lambda_0V({\bf p}_1,{\bf p}_2)$ is the pairing interaction.
The quasiparticle occupation numbers
\begin{equation}\label{3*}
n({\bf p})=v^2({\bf p})(1-f({\bf p}))+u^2({\bf p})f({\bf p}),
\end{equation}
and at finite temperatures the order parameter $\kappa$ reads
\begin{equation}\label{4*}
\kappa({\bf p})=v({\bf p})u({\bf p})(1-2f({\bf p})).
\end{equation}
While at $T=0$ the order parameter reduces to \cite{ksk}
\begin{equation}\label{17**}
\kappa({\bf p})=\sqrt{n_0({\bf p})(1-n_0({\bf p}))}.
\end{equation}
Here the coherence factors $v({\bf p})$ and $u({\bf p})$ are obeyed
the normalization condition
\begin{equation}\label{5*}
v^2({\bf p})+u^2({\bf p})=1.
\end{equation}
The distribution function $f({\bf p})$ of BQ defines the entropy
\begin{equation}\label{6*}
S=-2\int\left[f({\bf p})\ln f({\bf p})+(1-f({\bf p}))\ln(1-f({\bf
p}))\right]\frac{d{\bf p}}{4\pi^2}.
\end{equation}
We assume that the pairing interaction $\lambda_0V({\bf p}_1,{\bf
p}_2)$ is weak and produced, for instance, by electron---phonon
interaction. Minimizing $\Omega$ with respect to $\kappa({\bf p})$
and using the definition $\Delta({\bf p})=-\delta
\Omega/\kappa({\bf p})$, we obtain the equation connecting the
single-particle energy $\varepsilon({\bf p})$ to the
superconducting gap $\Delta({\bf p})$,
\begin{equation}\label{7*}
\varepsilon({\bf p})-\mu\ =\Delta({\bf p})\frac{1-2v^2({\bf p})}
{2v({\bf p})u({\bf p})}.
\end{equation}
The single-particle energy $\varepsilon({\bf p})$ is determined by
the Landau equation
\begin{equation}\label{8*}
\varepsilon({\bf p})= \frac{\delta E[n({\bf p})]}{\delta n({\bf
p})}.
\end{equation}
Note that $E[n({\bf p})]$, $\varepsilon[n({\bf p})]$, and the
Landau amplitude
\begin{equation}\label{9*}
F({\bf p},{\bf p}_1)=\frac{\delta E^2[n({\bf p})]}{\delta n({\bf
p})\delta({\bf p}_1)}
\end{equation}
implicitly depend on the number density $x$ which defines the
strength of $F$. Minimizing $\Omega$ with respect to $f({\bf p})$
and after some algebra, we obtain the equation for the
superconducting gap $\Delta({\bf p})$
\begin{equation}\label{10*}
\Delta({\bf p})=-\frac{1}{2}\int\lambda_0 V({\bf p},{\bf p}_1)
\frac{\Delta({\bf p}_1)}{E({\bf p}_1)}(1-2f({\bf p}_1))\frac{d{\bf
p}_1}{4\pi^2}.
\end{equation}
Here the excitation energy $E({\bf p})$ represented by BQ is given
by
\begin{equation}\label{11*}
E({\bf p})=\frac{\delta (E_{gs}-\mu N)}{\delta f({\bf p})}=
\sqrt{(\varepsilon({\bf p})-\mu)^2+\Delta^2({\bf p})}.
\end{equation}
The coherence factors  $v({\bf p})$, $u({\bf p})$, and the
distribution function $f({\bf p})$ are given by the ordinary
relations
\begin{equation}\label{12*}
v^2({\bf p})=\frac{1}{2}\left(1-\frac{\varepsilon({\bf
p})-\mu}{E({\bf p})}\right); u^2({\bf
p})=\frac{1}{2}\left(1+\frac{\varepsilon({\bf p})-\mu}{E({\bf
p})}\right),\end{equation}
\begin{equation}\label{13*}
f({\bf p})=\frac{1}{1+\exp(E({\bf p})/T)}.
\end{equation}
Equations \eqref{7*}---\eqref{13*} are the conventional equations
of the BCS theory \cite{bcs,til}, determining the superconducting
state with BQ and the maximum value of the superconducting gap
$\Delta_1\sim 10^{-3}\varepsilon_F$ provided that one assumes that
the system in question has not undergone FCQPT.

\begin{figure}[!ht]
\begin{center}
\includegraphics [width=0.47\textwidth]{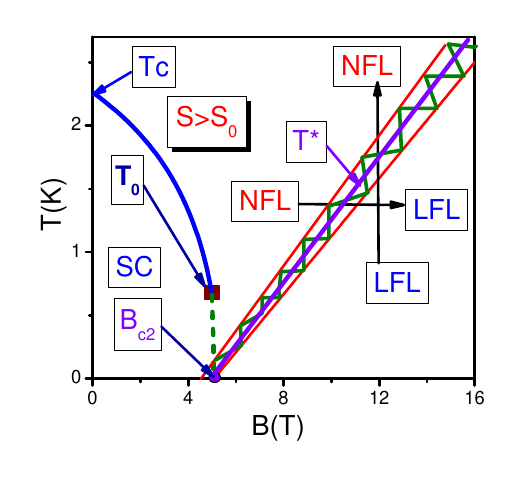}
%\vspace*{-0.4cm}
\end{center}%\vspace*{-0.8cm}
\caption {Schematic $T-B$ phase diagram of a superconducting HF
metal, with the upper critical field $B_{c2}$. The vertical and
horizontal arrows crossing the transition region are marked by the
line depicting the LFL-NFL and NFL-LFL transitions at fixed $B$ and
$T$, respectively. The hatched area indicates the crossover from
the LFL state, with $\rho(T)\propto T^2$, to the NFL one, with
$\rho(T)\propto T$. The median line $T^*$ of the crossover is shown
by the solid line. As shown by the solid curve, at $B<B_{c2}$ the
system is in its superconducting (SC) state. The superconducting
critical field $B_{c2}$ is shown by the violet circle.
Superconducting-normal phase boundary is displayed by the solid and
dashed curves. The solid square shows the point at $T=T_0$ where
the superconducting phase transition $T_c$ changes from the second
order phase transition to the first one.} \label{fig04}
\end{figure}
Now we consider a superconducting state with FC which takes place
after the FCQPT point. At $T=0$ and $\lambda_0\to 0$ the maximum
value of the superconducting gap $\Delta_1\to 0$, as well as the
critical temperature $T_c\to0$, and Eq. \eqref{7*} converts into
Eq. \eqref{7**} \cite{ks,ksk,phys_rep}. At $T\to0$, Eq. \eqref{7**}
defines the new state of Fermi liquid with FC \cite{ks,vol} which
is characterized by flat part of the spectrum in the $(p_f-p_i)$
region and has a strong impact on the system's properties up to
temperature $T_f^0$ \cite{ks,ms,Schuck,phys_rep}. It is seen from
Eq. \eqref{7**} that the entropy $S(T\to0)\to S_0$, where $S_0>0$
is given by
\begin{equation}\label{10**}
S_0=-\int[n_0(p)\ln n_0(p)+(1-n_0(p))\ln(1-n_0(p))]\frac{d{\bf
p}}{(2\pi)^2}.
\end{equation}
At $T\to0$, Eq. \eqref{7**} defines the particular state of a Fermi
liquid with FC, for which the modulus of the order parameter
$|\kappa({\bf p})|$ has finite values in the $(p_f-p_i)$ region,
whereas  $\Delta_1\to 0$ in this region. Observe that $f({\bf
p},T\to0)\to0$, and it follows from Eqs. \eqref{3*} and \eqref{4*}
that if $0<n({\bf p})<1$ then $|\kappa({\bf p})|\neq 0$ in the
region $(p_f-p_i)$. Such a state can be considered as
superconducting, with an infinitely small value of $\Delta_1$, so
that the entropy of this state is equal to zero. At any finite
$T>0$ the entropy $S\geq S_0$, thus the topological FQCPT is of the
first order \cite{phys_rep}. The FC state is formed by the Landau
interaction $F(p=p_F,p_1=p_F)$ being relatively strong as compared
with the pairing interaction $\lambda_0V$, therefore $\lambda_0V$
does not noticeably disturb the occupation numbers $n_0$, but does
disturb the corresponding flat band \cite{phys_rep,epl22}. If the
Landau interaction as a function of the number density $x$ is
sufficiently small, the flat part vanishes, and at $T\to0$ Eq.
\eqref{7**} has the only trivial solution $\varepsilon(p=p_F)=\mu$,
and the quasiparticle occupation numbers are given by the step
function, $n({\bf p})= \theta(p_F-p)$ \cite{ks,ksk}.

Consider the schematic phase diagram of unconventional high-$T_c$
superconductor. It is seen from the schematic phase diagram
\ref{fig04}, that at temperatures $T\lesssim T_c$ the
superconducting-normal phase transition shown by the solid line in
Fig. \ref{fig04} is of the second order and entropy $S$ is a
continuous function of its variable $T$ at $T_c(B)$. At
temperatures $T\to0$, the normal state can be recovered by the
application of magnetic field $B$, that is approximately equal to
the critical field $B\simeq B_{c2}$, and this state can be viewed
as the LFL one induced by the magnetic field. When the system in
its NFL state, under the application of magnetic field $B>T^*$, HF
metal transits to its LFL state, as seen from Fig. \ref{fig04}. At
$T\to0$ the entropy of the superconducting state $S_{SC}\to 0$ and
the entropy of the NFL state tends to some finite value
$S_{NFL}\geq S_0$, see Eq. \eqref{10**} \cite{ksk}. Thus, at
temperatures $T_0\geq T$ the equality $S_{SC}(T)=S_{NFL}(T)$ cannot
be satisfied \cite{phys_rep,shag06}. Thus, the second-order phase
transition becomes the first below a certain temperature
$T_{0}(B)$, as it happens in $\rm CeCoIn_5$ and as shown by the
arrow in Fig. \ref{fig04} \cite{shag06,bianchi,izawa}. We note that
the topological FCQPT is also of the first order. This first-order
phase transition is determined both by the entropy jump mentioned
above and by the topological charge of FCQPT, which also changes
abruptly \cite{vol,Volovik}. As a result, possible fluctuations of
the order parameter $\kappa$ are suppressed at $T\leq T_0$
\cite{phys_rep,shag06}.

\section{Common scaling of conventional and high-$T_c$
superconductors}\label{III}

\begin{figure}[!ht]
\begin{center}
%\vspace*{-0.8cm}
\includegraphics[width=0.5\textwidth]{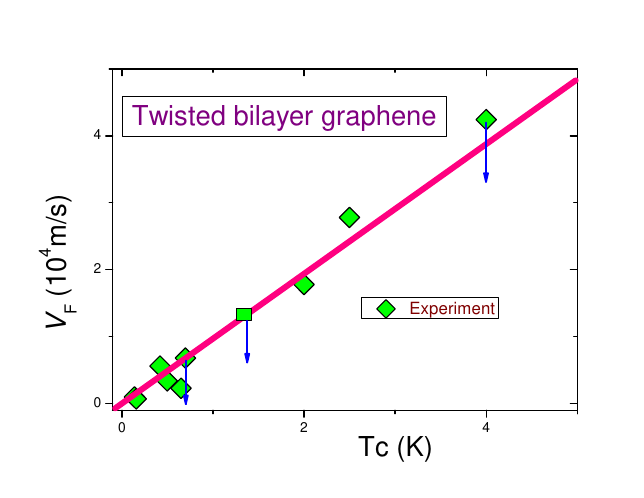}
\end{center}
%\vspace*{-0.8cm}
\caption{Experimental results for the average Fermi velocity $V_F$
as a function of the critical temperature $T_c$ for MATBG
\cite{mac}. The downward arrows show that $V_F\leq V_0$, with $V_0$
is the maximal shown value. Theory is shown by the solid line that
demonstrates $V_F\propto T_c\propto 1/N(0)$, see Eq. \eqref{15},
\cite{epl22}.} \label{fig05}
\end{figure}

It follows from Eqs. \eqref{7**} and \eqref{7*} that the system
becomes divided into two quasiparticle subsystems: the first
subsystem in the $(p_f-p_i)$ range is characterized  by the
quasiparticles with the effective mass $M^*_{FC}\propto
1/\Delta_1$, while the second one is occupied by quasiparticles
with finite mass $M^*_L$ and momenta $p<p_i$ \cite{phys_rep}. If
$\lambda_0\neq 0$, then $\Delta_1$ becomes finite. It is seen from
Eq. \eqref{10*} that the superconducting gap depends on the
single-particle spectrum $\varepsilon({\bf p})$. On the other hand,
it follows from Eq. \eqref{7*} that $\varepsilon({\bf p})$ depends
on $\Delta({\bf p})$, since at $\Delta_1\to0$ the spectrum becomes
flat. Let us assume that $\lambda_0$ is small so that the
particle-particle interaction $\lambda_0 V({\bf p},{\bf p}_1)$ can
only lead to a small perturbation of the order parameter
$\kappa({\bf p})$ determined by Eq. \eqref{17**}. Upon
differentiation both parts of Eq. \eqref{7*} with respect to the
momentum $p$, we obtain that the effective mass
$M^*_{FC}=d\varepsilon(p)/dp_{\,|p=p_F}$ becomes finite
\cite{epl22}
\begin{equation}\label{15*}
M^*_{FC}\sim p_F\frac{p_f-p_i}{2\Delta_1}.
\end{equation}
It follows from Eq. \eqref{15*} that the effective mass and the
density of states $N(0)\propto M^*_{FC}\propto 1/\Delta_1$ are
finite and constant at $T<T_c$ \cite{ms,epl22}. At $T\to0$ and
$\lambda_0\to 0$ the density of states near the Fermi level tends
to infinity. Thus, we arrive at the result that contradicts the BCS
theory, and follows from Eq. \eqref{15*}
\begin{equation}\label{15}
N(0)\propto M^*_{FC}\propto 1/\Delta_1\propto 1/T_c\propto 1/V_F,
\end{equation}
where $V_F\propto p_F/M^*_{FC}$ is the Fermi velocity
\cite{ms,mac}, see Fig. \ref{fig05}.

Measurements of $V_F$ as a function of $T_c$ \cite{mac} are
depicted in Fig. \ref{fig05}. These observations are in good a
agreement with Eq. \eqref{15}. Thus, our theoretical prediction
\cite{ms,phys_rep,shag} agrees very well with the experimental
results \cite{mac}. It is worth noting that $V_F\to 0$, as well as
$T_c\to 0$, as can be seen from Fig. \ref{fig05}. This result shows
that the flat band is disturbed by the finite value of $\Delta_1$,
and possesses a finite slope that makes $V_F\propto T_c$ , as seen
from Fig. \ref{fig05}. Indeed, from Fig. \ref{fig05}, the
experimental critical temperatures $T_c$ do not correspond to the
minima of the Fermi velocity $V_F$ as they would in any BCS-like
theory \cite{mac}. This extraordinary behavior is explained within
the framework of the FC theory based on the topological FCQPT,
forming flat bands \cite{phys_rep,epl22}. As we will see below,
another unusual behavior, i.e., the general universal scaling of
$E_{\Delta}/\gamma$ of both conventional and high-$T_c$
superconductors \cite{prb2015}, is also associated with Eq.
\eqref{15} and explained within the framework of the FC theory.

We are led to the conclusion that in contrast to the conventional
theory of superconductivity the single-particle spectrum
$\varepsilon({\bf p})$ strongly depends on the superconducting gap
and we have to solve Eqs. \eqref{8*} and \eqref{10*} in a
self-consistent way. On the other hand, let us assume that Eqs.
\eqref{8*} and \eqref{10*} are solved, and the effective mass
$M^*_{FC}$ is determined. Now one can fix the dispersion
$\varepsilon({\bf p})$ by choosing the effective mass $M^*$ of
system in question equal to $M^*_{FC}$ and then solve Eq.
\eqref{10*} as it is done in the case of the conventional theory of
superconductivity \cite{bcs}. As a result, one observes that the
superconducting state is characterized by BQ with the dispersion
given by Eq. \eqref{11*}, the coherence factors $v$, $u$ are given
by Eq. \eqref{12*}, and the normalization condition \eqref{5*} is
held. We conclude that the observed features agree with the
behavior of BQ in accordance with experimental facts
\cite{mat,nakam}. This observation suggests that the
superconducting state with FC is BCS-like and implies the basic
validity of BCS formalism in describing the superconducting state
in terms of BQ. It is exactly the case that was observed
experimentally in high-$T_c$ cuprates like
Bi$_2$Sr$_2$Ca$_2$Cu$_3$O$_{10+\delta}$, see e.g. \cite{mat,npbq}.

We now analyze other relationships  between the conventional
superconducting state and the superconducting state with FC. We
consider the case when $T_c\ll T_f^0$.  This means that the order
parameter $\kappa({\bf p})$ is slightly perturbed by the pairing
interaction because the particle-particle interaction $\lambda_0 V$
is small comparatively to the Landau amplitude $F$ and the order
parameter $\kappa({\bf p})$ is governed mainly by $F$
\cite{ks,phys_rep}. We can solve Eq. \eqref{10*} analytically
taking the BCS approximation for the particle-particle interaction:
$\lambda_0V({\bf p},{\bf p}_1)=\lambda_0$ if $|\varepsilon({\bf
p})-\mu|\leq \omega_D$, i.e. the interaction is zero outside this
region, with $\omega_D$ being the characteristic phonon energy. As
a result, the maximum value of the superconducting gap is given by
\cite{phys_rep}
\begin{equation}\label{16*}
\Delta_1\simeq \frac{\lambda_0 p_F(p_f-p_F)}{2\pi}
\ln\left(1+\sqrt2\right)
\end{equation}
$$\simeq 2\beta\varepsilon_F
\frac{p_f-p_F}{p_F}\ln\left(1+\sqrt2\right).$$ Here, the Fermi
energy $\varepsilon_F=p_F^2/2M^*_L$, and the dimensionless coupling
constant $\beta$ is given by the relation $\beta=\lambda_0
M^*_L/2\pi$. Taking the usual values of $\beta$ as $\beta\simeq
0.3$, and assuming $(p_f-p_F)/p_F\simeq 0.2$, we get from Eq.
\eqref{16*} a large value of $\Delta_1\sim 0.1\varepsilon_F$, while
for normal metals one has $\Delta_1\sim 10^{-3}\varepsilon_F$. Now
we determine the energy scale $E_0$ which defines the region
occupied by quasiparticles with the effective mass $M^*_{FC}$
\begin{equation}\label{17*} E_0=
\varepsilon({\bf p}_f)-\varepsilon({\bf p}_i) \simeq 2
\frac{(p_f-p_F)p_F}{M^*_{FC}}\ \simeq\ 2\Delta_1.
\end{equation}

We have returned back to the Landau Fermi liquid theory since the
high energy degrees of freedom are eliminated and the
quasiparticles are introduced. The only difference between LFL,
which serves as a basis when constructing the superconducting
state, and Fermi liquid with FC is that we have to expand the
number of relevant low energy degrees of freedom by introducing the
new type of quasiparticles with the effective mass $M^*_{FC}$ given
by Eq. \eqref{15*} and the energy scale $E_0$ given by Eq.
\eqref{17*}. Therefore, the dispersion $\varepsilon({\bf p})$ is
characterized by two effective masses $M^*_L$ and $M^*_{FC}$ and by
the scale $E_0$, which define the low temperature properties
including the line shape of quasiparticle excitations \cite{ms},
while the dispersion of BQ is given by Eq. \eqref{15}. We note that
both the effective mass $M^*_{FC}$ and the scale $E_0$ are
temperature independent at $T<T_c$, where $T_c$ is the critical
temperature of the superconducting phase transition
\cite{phys_rep}. Obviously, we cannot directly relate these new BQ
quasiparticle excitations with the quasiparticle excitations of an
ideal Fermi gas because the system in question has undergone the
topological FCQPT. However, the basic properties of the LFL theory
remains in FCQPT: low-energy excitations of a strongly correlated
liquid with FC are quasiparticles, whereas in the superconducting
state they are represented by BQ \cite{phys_rep}. As it was shown
above, properties of these new quasiparticles are closely related
to the properties of the superconducting state with the order
parameter at $T=0$ being given by Eq. \eqref{17**} We may say that
the quasiparticle system in the range $(p_f-p_i)$ becomes very
``soft', that is it should be adjusted to the superconducting
state, see Eqs. \eqref{15*} and Eq. \eqref{15}, and is to be
considered as a strongly correlated liquid.

\begin{figure}
\begin{center}
%\vspace*{-2.5cm}
\includegraphics [width=0.47\textwidth]{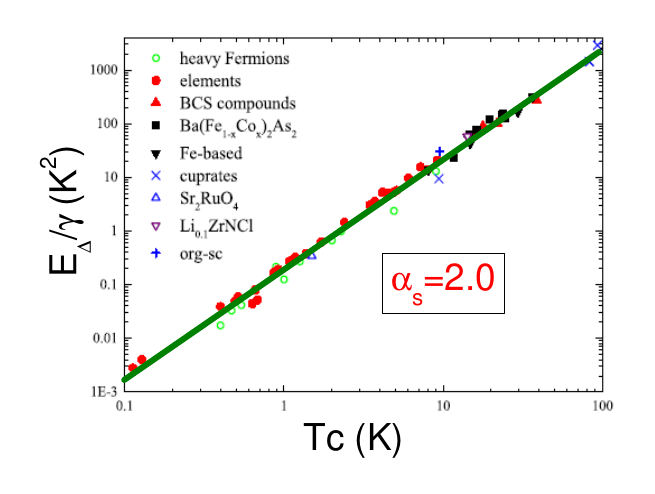}
%\vspace*{-0.75cm}
\end{center}
\caption{Condensation energy $E_\Delta/\gamma\propto T_c^{2}$
divided by the specific heat $\gamma$ as a function of $T_c$ for a
wide range of superconductors, with the slope $\alpha_{s}=2$
\cite{prb2015}, see Eq. \eqref{30}. Deviations from the line of
best fit, spanning six orders of magnitude for $E_\Delta/\gamma$
and almost three orders of magnitude for $T_c$, are relatively
small.}\label{fig1}
\end{figure}
At the same time, one could expect serious deviations from the BCS
results when calculating the pairing correction $\Delta E_{FC}$ to
$E_0[n]$. Applying the Landau formula for the change of $E_0[n]$
due to the variation $\delta n({\bf p},T)=n({\bf p},T)-n_0({\bf
p})$ of the occupation numbers \cite{lanl} and adding the
superfluid term \eqref{2**} we arrive at the following result
\begin{equation}\label{28**}
\Delta E_{FC}=\int(\varepsilon({\bf p})-\mu)\delta n({\bf
p}){d^3p\over(2\pi)^3}+\delta E_s.
\end{equation}
Here $\delta E_s$ is given by Eqs. \eqref{TC}, \eqref{2**} and
\eqref{17**}
\begin{equation}\label{29**}
\delta E_{s}=-{1\over 2}\int_{p_i}^{p_f} \Delta_0(p)\sqrt
{n_0(p)(1-n_0(p))}{p^2dp\over 2\pi^2}.
\end{equation}
In the usual BCS case the first term and the second one become
proportional $\sim\Delta^2/\varepsilon^0_F$ so that $\Delta
E_{FC}\sim\Delta^2/\varepsilon^0_F$ \cite{lanl}. One could suspect
that in the system with the FC the first term in Eq. \eqref{28**}
turns out to be zero, for $\varepsilon({\bf p})-\mu=0$ in the
region $p_f-p_i$, see Eq. \eqref{7**}. This is not true, since both
the Fermi velocity $V_F$ and the effective mass $M_{FC}^*$ become
finite under the influence of the superconducting state
\cite{epl22}, see Eq. \eqref{15}. Considering also that we are
dealing with BQ, we are left with the usual BCS result for the
superconducting condensation energy $E_{\Delta}$, which is valid
for both conventional superconductors and unconventional
superconductors with high-$T_c$,
\begin{equation}\label{30}
\Delta E_{FC}/\gamma\sim E_{\Delta}/\gamma\sim
\frac{N(T)\Delta_1^2}{\gamma(T)}\sim \Delta_1^2\sim T_c^2.
\end{equation}
Here $N(T)$ and $\gamma(T)$ are the density of states and the
Sommerfeld coefficient, correspondingly. $N(T)$ and $\gamma(T)$
strongly depend on temperature $T$ in the FC theory, and $\Delta_1$
is the maximum value of the superconducting gap. However,
$M^*(T)\propto N(T)\propto \gamma(T)$ \cite{phys_rep}, and we
obtain $E_{\Delta}/\gamma\sim T_c^2$. It is seen from Fig.
\ref{fig1} that Eq. \eqref{30} is in accordance with experimental
facts \cite{prb2015}. Indeed, taking into account that BQ of
unconventional high-$T_c$ superconductors within the framework of
the FC theory coincide with BQ of conventional superconductors and
Eq. \eqref{15*}, we conclude that the condensation energy
$E_{\Delta}/\gamma$ given by Eq. \eqref{30} has the universal form
valid in the case of both conventional superconductors and
high-$T_c$ ones. To check this conclusion, we compare our
theoretical result with experimental facts \cite{prb2015}. Figure
\ref{fig1} shows the scaling of the condensation energy
$E_{\Delta}$ versus $T_c^2$ on log-log scale. It is seen from Fig.
\ref{fig1} that the universal scaling $E_{\Delta}/\gamma\propto
T_c^2$ is valid for all superconductors, both the conventional and
the unconventional high-$T_c$ ones. This universal scaling behavior
takes place over almost seven orders of magnitude for
$E_{\Delta}/\gamma$ and three orders of magnitude for $T_c$
\cite{prb2015}. This observation is not surprising, for, as we have
seen above, high-$T_c$ superconductors have the same BQ as
conventional one, since the shape of the corresponding band is
correlated with their $T_c$, as it follows from Eq. \eqref{15}.
Note that due to the strong influence of the pseudogap state on the
properties of unconventional superconductors, such as the density
of states, heat capacity, and even the real meaning of $T_c$ is not
clear, only optimally doped samples were considered
\cite{prb2015,loram}. Thus, the FC theory allows one to justify Eq.
\eqref{30} that describes the superconductivity extending far
beyond the weak coupling regime, and applies to both the
conventional and the unconventional strongly correlated
superconductors.

\section{Summary}\label{IV}
We have analyzed the common behavior of unconventional high-$T_c$
and conventional superconductors and demonstrated that the
universal scaling of the condensation energy
$E_{\Delta}/\gamma=N(0)\Delta_1^2/\gamma$  applies equally to
conventional and unconventional high-$T_c$ superconductors. Our
explanation is based on the general property of superconductors:
Bogoliubov quasiparticles act in conventional and unconventional
superconductors, while the corresponding band is only deformed by
the unconventional superconducting state. These observations
suggest that the unconventional superconducting state can be
considered BCS-like in some cases, as predicted in
\cite{phys_rep,ms,jetplbq}. Our theoretical observations are in
good agreement with experimental facts.

%\section{Acknowledgement}
We thank V.A. Khodel for fruitful discussions. This work was
supported by U.S. DOE, Division of Chemical Sciences, Office of
Basic Energy Sciences, Office of Energy Research, AFOSR.


\begin{thebibliography}{99}

\bibitem{catal} N. Regnault, Y. Xu, M.-R. Li, D.-Sh. Ma, M.
Jovanovic {\it et al.}, Nature {\bf 603}, 824 (2022).

\bibitem{scn} L. Chen, D. T. Lowder, E. Bakali, A. M. Andrews, W. Schrenk, et
al., Science {\bf 382}, 907 (2023).

\bibitem{mat} H. Matsui, T. Sato, T. Takahashi, S.-C. Wang, H.-B. Yang,
H. Ding, T. Fujii, T. Watanabe, and A. Matsuda, Phys. Rev. Lett.
{\bf 90}, 217002 (2003).

\bibitem{prb2015} J.S. Kim, G.N. Tam, and G.R. Stewart, Phys. Rev. B {\bf 92},
224509 (2015).

\bibitem{prlq} A. Hunter, S. Beck, E. Cappelli, F. Margot, M. Straub {\it et al}.,
Phys. Rev. Lett. {\bf 131}, 236502 (2023).

\bibitem{npbq} K.-J. Xu, Qinda Guo, M. Hashimoto, Z.-X. Li, S.-D. Chen
{\it et al}., Nat. Phys. {\bf 19}, 1834 (2023).

\bibitem{mac} W. Qin, B. Zou, and A.H. MacDonald, Phys. Rev. B
{\bf 107}, 024509 (2023).

\bibitem{epl22} V.R. Shaginyan, A.Z. Msezane, M.Ya. Amusia,
and G.S. Japaridze, EPL, {\bf 138}, 16004 (2022).

\bibitem{lanl} E. M. Lifshitz, L.~Pitaevskii,
\emph{Statistical Physics. Part 2.} (Butterworth-Heinemann, Oxford,
2002).

\bibitem{ks} V.A. Khodel and V.R. Shaginyan, JETP Lett. {\bf 51}, 553 (1990).

\bibitem{ksk} V.A. Khodel, V.R. Shaginyan, and V.V. Khodel, Phys. Rep. {\bf
249}, 1 (1994).

\bibitem{vol} G.E. Volovik, JETP Lett. {\bf 53}, 222
(1991).

\bibitem{phys_rep} V.R. Shaginyan, M.Ya. Amusia, A.Z. Msezane,
and K.G. Popov, Phys. Rep. {\bf 492}, 31 (2010).

\bibitem{Volovik} T.T. Heikkila and G.E. Volovik, Flat bands as a route
to high-temperature superconductivity in graphite. Springer Series
in Materials Science, Vol. {\bf 244} (Springer Nature Switzerland
AG, Cham, 2016).

\bibitem{book_20} M.Ya. Amusia and V.R. Shaginyan,
{\it Strongly Correlated Fermi Systems: A New State of Matter},
Springer Tracts in Modern Physics Vol. {\bf 283} (Springer Nature
Switzerland AG, Cham, 2020).

\bibitem{prl20} P. Rosenzweig, H. Karakachian, D. Marchenko, and
K. K\"uster, and U. Starke, Phys. Rev. Lett. {\bf 125}, 176403
(2020).

\bibitem{bern} P. T\"orm\"a,  S. Peotta, and B.A. Bernevig,
Nat. Rev. Phys. {\bf 4}, 528 (2022).

\bibitem{bern_prl} V. Peri, Z.D. Song,
B.A. Bernevig, and S.D. Huber, Phys. Rev. Lett. {\bf 126}, 027002
(2021).

\bibitem{bcs} J. Bardeen, L.N. Cooper, and J.R. Schriffer, Phys.
Rev. {\bf 108}, 1175 (1957).

\bibitem{til} K.H. Bennemann and J.B. Ketterson, {\it
Superconductivity}, (Springer-Verlag Berlin Heidelberg, 2008).

\bibitem{jetplbq} M.Ya. Amusia and V.R. Shaginyan, JETP Lett. {\bf 77},
671 (2003).

\bibitem{shag} V.R. Shaginyan, JETP Lett. {\bf 77}, 99
(2003).

\bibitem{Schuck} J. Dukelsky, V.A. Khodel, P. Schuck, and V.R.
Shaginyan, Z. Phys. B{\bf 102}, 245 (1997).

\bibitem{ms} M.Ya. Amusia and V.R. Shaginyan,
Phys. Rev. B {\bf 63}, 224507 (2001).

\bibitem{pla98} V.R. Shaginyan, Phys. Lett. A{\bf 249}, 237 (1998).

\bibitem {shag06} V.R. Shaginyan, A.Z. Msezane, V.A. Stephanovich, and E.V.
Kirichenko, EPL {\bf 76}, 898 (2006).

\bibitem{bianchi} A. Bianchi, R. Movshovich, N. Oeschler,
P. Gegenwart, F. Steglich {\it et al.}, Phys. Rev. Lett. {\bf 89},
137002 (2002).

\bibitem{izawa} K. Izawa, H. Yamaguchi,
Y. Matsuda, H. Shishido, R. Settai, and Y. Onuki, Phys. Rev. Lett.
{\bf 87}, 057002 (2001).

\bibitem{nakam}  S. Nakamae, K. Behnia, N. Mangkorntong, M. Nohara,
H. Takagi, S. J. C. Yates, and N. E. Hussey,  Phys. Rev. B {\bf
68}, 100502(R) (2003).

\bibitem{loram} J.W. Loram, K.A. Mirza, J.M. Wade, J.R. Cooper, and W.Y.
Liang, Physica C {\bf 235-240}, 134 (1994).

\end{thebibliography}
\end{document}